\documentclass[conference]{IEEEtran}

\IEEEoverridecommandlockouts

\usepackage[T1]{fontenc}
\usepackage[utf8]{inputenc}
\usepackage{cite}
\usepackage{amsmath,amssymb,amsfonts}
\usepackage{algorithmic}
\usepackage{graphicx}
\usepackage{textcomp}
\usepackage{xcolor}
\definecolor{ForestGreen}{RGB}{34,139,34}
\definecolor{orange}{RGB}{255,165,0}
\usepackage{pifont}
\usepackage{listings}
\usepackage{enumitem}
\usepackage{multirow}
\usepackage{algorithm}
\usepackage{booktabs}
\usepackage{array}
\usepackage{tabularx}
\usepackage{makecell}
\usepackage{colortbl}
\usepackage{newunicodechar}
\usepackage{comment}
\usepackage{csquotes}
\usepackage{url}

\newunicodechar{σ}{$\sigma$}
\newunicodechar{Δ}{$\Delta$}
\newunicodechar{−}{\ensuremath{-}} 
\newunicodechar{→}{\ensuremath{\to}} 
\newunicodechar{–}{--} 
\newunicodechar{—}{---} 
\newunicodechar{“}{``} 
\newunicodechar{”}{''} 

\usepackage{stfloats}

\usepackage{glossaries}
\newacronym{os}{OS}{Operating System}
\newacronym{nyi}{NYI}{Not Yet Implemented}
\newacronym{AI}{AI}{Artificial Intelligence}
\newacronym{BLEU}{BLEU}{Bilingual Evaluation Understudy}
\newacronym{ROUGE}{ROUGE}{Recall-Oriented Understudy for Gisting Evaluation}
\newacronym{METEOR}{METEOR}{Metric for Evaluation of Translation with Explicit ORdering}
\newacronym{ML}{ML}{Machine Learning}
\newacronym{NLP}{NLP}{Natural Language Processing}
\newacronym{ICL}{ICL}{In-Context Learning}
\newacronym{RAG}{RAG}{Retrieval Augmented Generation}
\newacronym{LSTM}{LSTM}{Long Short-Term Memory}
\newacronym{RNNs}{RNNs}{Recurrent Neural Networks}
\newacronym{BERT}{BERT}{Bidirectional Encoder Representations from Transformers}
\newacronym{MLM}{MLM}{Masked Language Modeling}
\newacronym{NSP}{NSP}{Next Sentence Prediction}
\newacronym{ASTs}{ASTs}{Abstract Syntax Trees}
\newacronym{MBPP}{MBPP}{Mostly Basic Python Problems}
\newacronym{RLHF}{RLHF}{Reinforcement Learning from Human Feedback}
\newacronym{GenAI}{GenAI}{Generative Artificial Intelligence}
\newacronym{CoT}{CoT}{Chain-of-Thought}
\newacronym{LLOC}{LLOC}{Logical Lines of Code}
\newacronym{SLOC}{SLOC}{Source Lines of Code}
\newacronym{LOC}{LOC}{Lines of Code}
\newacronym{MoE}{MoE}{Mixture of Experts}
\newacronym{GMM}{GMM}{Gaussian Mixture Models}
\newacronym{SPV}{SPV}{Software Patch Verification}
\newacronym{VD}{VD}{Vulnerability Detection}
\newacronym{APR}{APR}{Automated Program Repair}
\newacronym{PBD}{PBD}{Primary Benchmark Dataset}
\newacronym{LFD}{LFD}{Leakage Free Dataset}
\newacronym{PPR}{PPR}{Positive Prediction Rate}
\newacronym{TP}{TP}{True Positive}
\newacronym{FP}{FP}{False Positive}
\newacronym{TN}{TN}{True Negative}
\newacronym{FN}{FN}{False Negative}
\newacronym{MRR}{MRR}{Mean Reciprocal Rank}
\newacronym{HDD}{HDD}{Hierarchical Distance and Direction}
\newacronym{DFI}{DFI}{Directional Failure Index}
\newacronym{CWE}{CWE}{Common Weakness Enumeration}
\newacronym{CVE}{CVE}{Common Vulnerabilities and Exposures}
\newglossaryentry{SAT}
{
  name={SAT},
  description={Static Analysis Tool},
  first={Static Analysis Tool (SAT)},
  plural={SATs},
  descriptionplural={Static Analysis Tools},
  firstplural={Static Analysis Tools (SATs)}
}

\newglossaryentry{LLM}
{
  name={LLM},
  description={Large Language Model},
  first={Large Language Model (LLM)},
  plural={LLMs},
  descriptionplural={Large Language Models},
  firstplural={Large Language Models (LLMs)}
}

\newacronym{ODC}{ODC}{Orthogonal defect classification}

\usepackage[hidelinks]{hyperref}

\lstdefinelanguage{Diff}{
    language=C,
    morecomment=[f][\color{red}]{-},
    morecomment=[f][\color{ForestGreen}]{+},
}

\lstdefinestyle{customc}{
  belowcaptionskip=1\baselineskip,
  breaklines=true,
  frame=L,
  xleftmargin=\parindent,
  language=C,
  showstringspaces=false,
  numbers=left,               
  numberstyle=\tiny\color{gray}, 
  stepnumber=1,
  basicstyle=\footnotesize\ttfamily, 
  keywordstyle=\bfseries\color{green!40!black},
  commentstyle=\itshape\color{purple!40!black},
  numbersep=8pt,
  identifierstyle=\color{blue},
  stringstyle=\color{orange},
}

\definecolor{diffgreen}{rgb}{0.0, 0.5, 0.0} 
\definecolor{diffred}{rgb}{0.6, 0.0, 0.0}   
\definecolor{codedark}{rgb}{0.2, 0.2, 0.2}  
\definecolor{codegray}{rgb}{0.5, 0.5, 0.5}  
\definecolor{codeblue}{rgb}{0.0, 0.0, 0.8}  

\lstdefinelanguage{customdiff}{
  language=C,
  morecomment=[f][\color{diffred}]{-},          
  morecomment=[f][\color{diffgreen}]{+},        
  morecomment=[f][\color{codeblue}]{@@},        
  morecomment=[l][\color{codegray}]{//},        
  moredelim=**[is][\bfseries\color{black}]{@}{@}, 
}

\definecolor{Gray}{gray}{0.92}

\def\BibTeX{{\rm B\kern-.05em{\sc i\kern-.025em b}\kern-.08em
    T\kern-.1667em\lower.7ex\hbox{E}\kern-.125emX}}

\title{Calibration Without Comprehension: Diagnosing the Limits of Fine-Tuning LLMs for Vulnerability Detection in Systems Software}

\author{
  \IEEEauthorblockN{Arastoo Zibaeirad,\; Marco Vieira}
  \IEEEauthorblockA{University of North Carolina at Charlotte, Charlotte, NC, USA\\
    Email: \{azibaeir, marco.vieira\}@charlotte.edu\\
    \includegraphics[height=1.8ex]{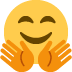}\ \href{https://huggingface.co/Arastoorad/CWE_Trace-lora-adapters}{Hugging Face}
    \hspace{1.2em}
    \includegraphics[height=1.8ex]{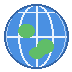}\ \href{https://erroristotle.github.io/projects/cwe-trace/}{Website}}
}

\newlength{\mediumrulewidth}
\newcolumntype{C}[1]{>{\centering\arraybackslash}p{#1}}

\begin{document}
\bstctlcite{IEEEexample:BSTcontrol}
\maketitle

\begin{abstract}
    Whether \glspl{LLM} scoring well on vulnerability benchmarks genuinely reason about security or merely pattern-match on contaminated data remains unresolved. We present \textbf{CWE-Trace}, a framework for \gls{LLM} vulnerability detection built from 834 manually curated Linux kernel samples spanning 74 \glspl{CWE}. The framework enforces a strict temporal split (pre-2025 historical set / post-cutoff leakage-free set), preserves context-aware vulnerable--patched pairs, and introduces two diagnostic metrics: the Directional Failure Index (DFI) and Hierarchical Distance and Direction (HDD). We evaluate eight vanilla \glspl{LLM} and 15 LoRA fine-tuned variants across non-targeted detection, targeted detection, and \gls{CWE} classification. Our analysis yields two key results. First, \textbf{data contamination provides no measurable advantage.} Function-level analysis shows that 84\% of nominally contaminated samples carry no usable memorization signal: vulnerable functions are absent or cross-mapped across datasets, and ${\sim}31\%$ of contaminated samples carry CWE misclassification. Second, \textbf{backbone directional priors dominate fine-tuning.} Models exhibit stable, systematic failure modes (DFI ranging from $-85.5$ to $+94.8$~pp) that persist from historical to post-cutoff data and resist correction. Fine-tuning shifts the output threshold without changing the decision policy. This is calibration without comprehension: output distributions adapt to training data while the underlying security reasoning remains absent. The weakest backbone at binary detection (DeepSeek-R1) gains the most in coarse \gls{CWE} classification, revealing that detection and understanding are decoupled capabilities. The best detection score reaches only $52.1\%$ ($+2.1$~pp above chance); exact \gls{CWE} ranking remains below $1.3\%$ Top-1 accuracy, confirming that current \glspl{LLM} lack reliable security reasoning for systems software, regardless of fine-tuning strategy.
\end{abstract}

\glsresetall

\begin{IEEEkeywords}
    Large Language Models, Vulnerability Detection, CWE Classification, Supervised Fine-tuning
\end{IEEEkeywords}

\section{Introduction}

Software vulnerabilities threaten modern systems at scale, with reported \glspl{CVE} exceeding 29,000 in 2023 alone~\cite{cvedetails2024}, yet automated detection tools remain far from reliably deployable. Systems software such as the Linux kernel is a particularly demanding target: low-level C code, memory-unsafe primitives, and cross-file dependencies make vulnerabilities both hard to detect and consequential when missed. Real-world flaws rarely appear as isolated snippets; they emerge from interactions across functions and files. Effective detectors must flag vulnerable code and identify the underlying \gls{CWE} to support triage and mitigation.

\begin{figure}
    \centering
    \includegraphics[width=1\columnwidth]{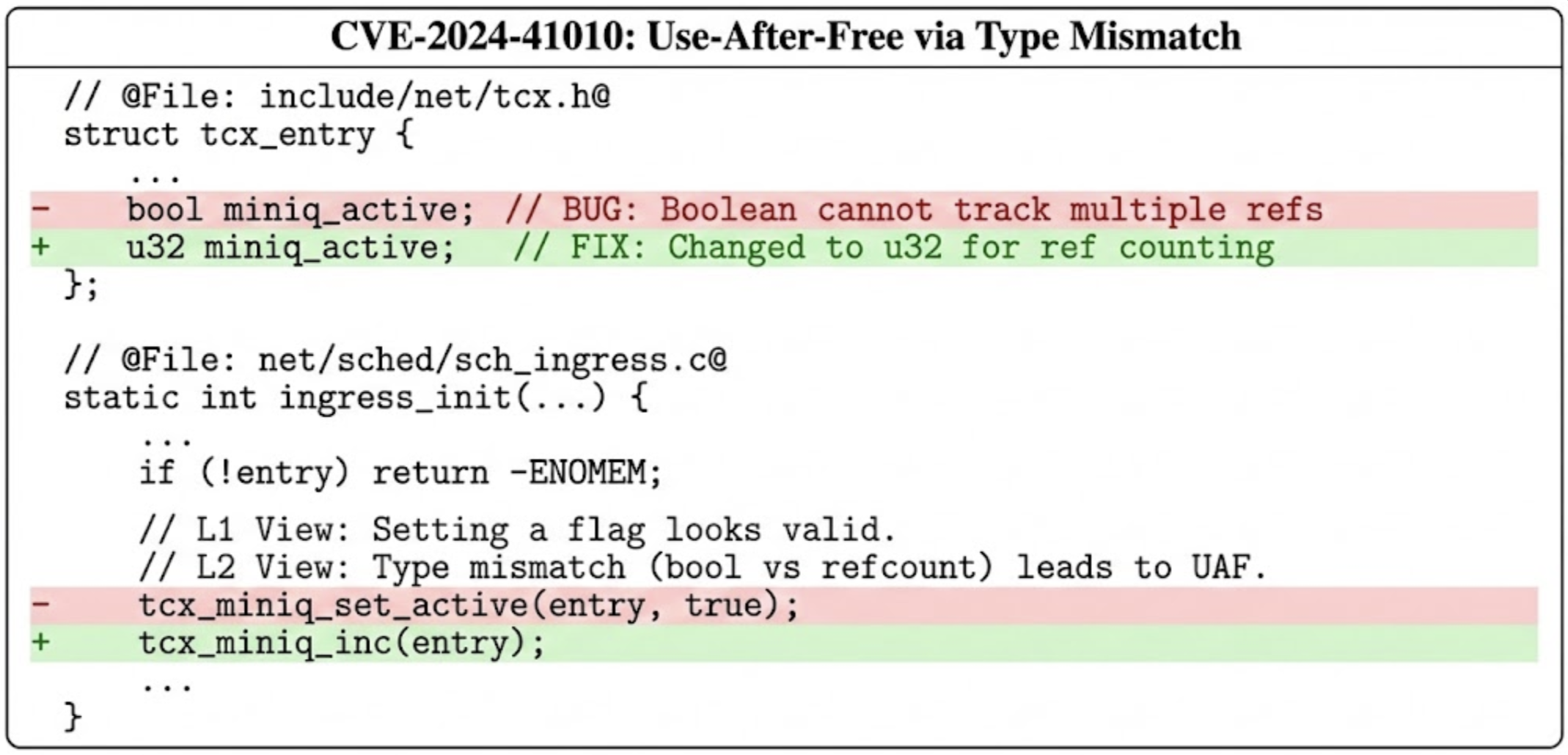}
    \vspace{-12pt}
    \caption{A unified diff of CVE-2024-41010. The vulnerability (Use-After-Free) is invisible in \texttt{sch\_ingress.c} unless the model also verifies the type definition in \texttt{tcx.h}, highlighting the necessity of L2 (multi-file) context.}
    \label{lst:cve-2024-41010}
    \vspace{-12pt}
\end{figure}

Traditional approaches (e.g., fuzzing, penetration testing, and \glspl{SAT} such as CodeQL~\cite{codeql} and Semgrep~\cite{semgrep}) suffer from coverage gaps, logic blindness, and false positives~\cite{pereira2021machine}. Recent work explores \glspl{LLM} for code and security tasks, including vulnerability benchmarks and CWE analyses~\cite{chen2021evaluating, touvron2023llama, feng2020codebert, li2022competition, siddiq2022securityeval, liu2024vuldetectbench, sun2024llm4vuln, khare2025understanding, ullah2024llms, ghorbani2025examining, rashid2024quantifying}.
However, evaluations often (1) rely on historical vulnerabilities with possible contamination~\cite{ullah2024llms}, (2) strip context, (3) lack strict vulnerable--patched pairing~\cite{gao2023far, iannone2022secret, croft2023data}, and (4) report only binary accuracy, obscuring hierarchical \gls{CWE} understanding.
Gap~(2) is illustrated by CVE-2024-41010 (Figure~\ref{lst:cve-2024-41010}): the Use-After-Free (CWE-416) is invisible when \texttt{ingress\_init} is analyzed alone and revealed only by checking \texttt{tcx.h} (a \texttt{bool} flag). These gaps prevent a reliable assessment of whether fine-tuning \glspl{LLM} instills genuine security reasoning or merely \emph{calibrates without comprehension}: shifting the output distribution to match training labels without improving the underlying representations that support security reasoning.

Prior work on vulnerability detection mainly studies either non-\gls{LLM} models (e.g., GNNs, short-context Transformers) or applies \glspl{LLM} to short, single-function snippets, without jointly addressing long-context vulnerabilities and fine-tuning long-context \glspl{LLM} across multiple state-of-the-art VD datasets. For example, Ullah et al.~\cite{ullah2024llms} highlight contamination risks using a curated set of 228 samples across 8 CWEs, but do not examine function-level distribution shifts, multi-backbone fine-tuning, or diagnostic decomposition. Similarly, Croft et al.~\cite{croft2023data} identify label noise at the dataset level. In contrast, we quantify compromised-sample rates at the function level and link them mechanistically to a null contamination finding.

We present CWE-Trace, a framework that addresses limitations through design principles. First, we introduce a \emph{temporal split}: a \gls{PBD} covers vulnerabilities prior to 2025, while a \gls{LFD} includes only CVEs from 2025 onward, enabling evaluation of generalization vs.\ memorization. Second, we use a context-aware approach aligning 417 Linux kernel CVEs with vulnerable--patched code pairs, removing non-security changes and retaining functions when vulnerabilities span multiple interactions. Third, we evaluate models in non-targeted and targeted settings, measuring binary detection and CWE classification via \gls{DFI} and \gls{HDD}, two metrics that characterize \emph{how} models fail beyond aggregate accuracy.

We evaluate eight vanilla \glspl{LLM} spanning code-specialized and general-purpose across 74 \gls{CWE} types, and fine-tune three backbones (Qwen3{-}4B, DeepSeek-R1{-}32B, and Llama3.1{-}8B) on five SOTA vulnerability datasets: Devign~\cite{zhou2019devign}, LineVul~\cite{fu2022linevul}, PrimeVul~\cite{ding2024vulnerability}, MegaVul~\cite{ni2024megavul}, and VDISC~\cite{russell2018automated}. These backbones span scales (4B, 8B, 32B), families, and pre-training distributions, while tractable for LoRA fine-tuning. This yields 15 supervised LoRA baselines. Across both settings, our results reveal calibration without comprehension: fine-tuning adjusts outputs without instilling security reasoning, leaving detection near chance and CWE understanding below operational utility.

In summary, the contributions of this work are:
\begin{itemize}
    \item \textbf{Benchmark and dataset.} 834 context-aware vulnerable--patched samples from 417 Linux kernel CVEs, verified through a two-reviewer inspection protocol. Unlike prior contamination-aware benchmarks~\cite{ullah2024llms}, we perform function-level distribution-shift verification to distinguish CVE-level overlap from actual memorization exposure.

    \item \textbf{Systematic fine-tuning study.} 15 LoRA variants (3 backbones $\times$ 5 SOTA datasets) evaluated on the same benchmark, with contamination controls and label-space analysis isolating what fine-tuning actually changes versus what remains determined by backbone priors.

    \item \textbf{Diagnostic metrics.} \gls{DFI} and \gls{HDD} characterize \emph{how} models fail beyond aggregate accuracy: DFI decomposes near-random scores into directional bias; HDD measures hierarchical error structure in CWE-1000.

    \item \textbf{Data quality audit.} Function-level verification of contaminated CVEs across PrimeVul, MegaVul, and LineVul, quantifying semantic conflict, CWE misclassification, and distribution shift in widely-used corpora; the null result is mechanistically explained via distribution-shift analysis (more precise than dataset-level audits~\cite{croft2023data}).

\end{itemize}

The paper is structured as follows: \S\ref{section:related_work} reviews related work; \S\ref{section:approach} presents methodology; \S\ref{section:setup} details setup; \S\ref{section:rq1}--\S\ref{section:rq3} analyze results; \S\ref{section:discussion} discusses implications; \S\ref{section:threatstovalidity} states threats; and \S\ref{section:conclusion} concludes.

\section{Background and Related Work}
\label{section:related_work}

\begin{table}[!t]
    \centering
    \caption{Vulnerability detection datasets at a glance.}
    \label{tab:bench-compare}
    \small
    \setlength{\tabcolsep}{3pt}
    \renewcommand{\arraystretch}{1.1}
    \resizebox{\columnwidth}{!}{%
        \begin{tabular}{lccccc}
            \toprule
            \textbf{Dataset}                     & \textbf{Scale} & \textbf{CWE} & \textbf{Pairs (vuln vs patch)} & \textbf{Leakage} & \textbf{Granularity} \\
            \midrule
            \cite{lu2021codexglue} (2021)        & 27{,}318       & --           & \ding{55}                      & \ding{55}        & Func                 \\
            \cite{siddiq2022securityeval} (2022) & 130            & 75           & \ding{51}                      & \ding{55}        & Snippet              \\
            \cite{liu2024vuldetectbench} (2024)  & 1000           & 48           & \ding{51}                      & \ding{55}        & Program              \\
            \cite{ullah2024llms} (2024)          & 228            & 8            & \ding{51}                      & \ding{51}        & Func.                \\
            \cite{khare2025understanding} (2025) & 5{,}000        & 25           & partial                        & \ding{55}        & Func.+project        \\
            CWE-Trace                            & 834            & 74           & \ding{51}                      & \ding{51}        & Func.+context        \\
            \bottomrule
        \end{tabular}%
    }
    \vspace{-12pt}
\end{table}

\paragraph{Vulnerability Datasets (Granularity and Context)}
Early research prioritized scale through \emph{file-level} labeling, marking entire files as vulnerable based on commit logs~\cite{chakraborty2021deep, zheng2021d2a, ni2024megavul}. That scale introduced noise by conflating security fixes with unrelated edits~\cite{gao2023far, iannone2022secret}. Subsequent work shifted to \emph{function-level} granularity~\cite{zhou2019devign, bhandari2021cvefixes}, yet often extracted functions in isolation, severing the cross-component dependencies that define complex flaws. In contrast, CWE-Trace adopts a \emph{context-aware} approach: we pair strictly validated vulnerable and patched code blocks while preserving the surrounding non-functional elements (macros, type definitions) essential for semantic interpretation, moving beyond isolated snippets to capture the complete multi-file patch context.

\paragraph{LLM Evaluation (Contamination and Rigor)}
\gls{LLM} evaluation has evolved from general code tasks~\cite{lu2021codexglue} to security-specific benchmarks~\cite{siddiq2022securityeval, liu2024vuldetectbench, sun2024llm4vuln}, but a critical validity threat remains: \emph{data contamination}. Most benchmarks rely on historical CVEs likely present in training corpora, making genuine reasoning indistinguishable from memorization. The closest prior work~\cite{ullah2024llms} acknowledges this with a leakage-aware benchmark, but covers only 228 samples across 8 CWEs, and neither conducts fine-tuning experiments nor performs function-level distribution-shift verification. CWE-Trace addresses these gaps with a larger and more diverse benchmark, a multi-backbone LoRA study, a strict temporal split that tests generalization to unseen threats, and a mechanistic explanation for why CVE-level overlap does not entail memorization.

\paragraph{Taxonomy and Hierarchical Reasoning}
The \gls{CWE} hierarchy~\cite{mitre_cwe} provides a structured taxonomy of weaknesses, yet most evaluations flatten it into binary or multi-class labels. That loses partial understanding, where a broad but related class is better than an unrelated guess. By integrating hierarchical analysis into evaluation, we distinguish models that grasp the nature of a flaw from those that merely hallucinate unrelated labels. This distinction is important for security triage, where a related family prediction can still narrow investigation.

\paragraph{Supervised Fine-Tuning for Vulnerability Detection}
Prior work fine-tunes Transformer or GNN architectures on vulnerability datasets~\cite{zhou2019devign, fu2022linevul, ni2024megavul} or applies instruction-tuned \glspl{LLM} to security tasks~\cite{sun2024llm4vuln, khare2025understanding}, but typically evaluates a single architecture on its own training corpus, conflating backbone capability with dataset effects. No prior work systematically evaluates multiple \gls{LLM} backbones across multiple \gls{VD} datasets under matched contamination controls, the design needed to isolate genuine security learning from training-distribution artifacts.


\begin{figure*}[!t]
    \centerline{\includegraphics[width=1\textwidth]{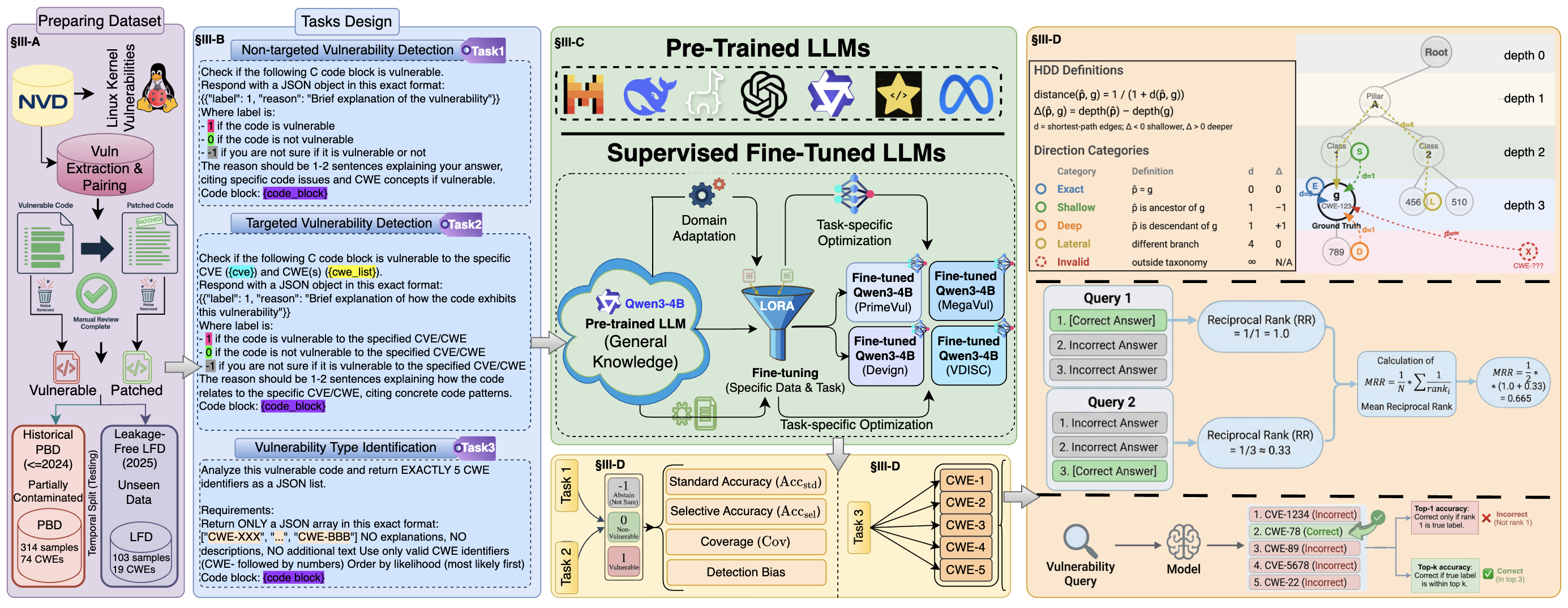}}
    \vspace{-3pt}
    \caption{Overview of the CWE-Trace framework. The pipeline begins with the extraction and manual pairing of 834 Linux kernel samples (417 pairs) (\S\ref{section:dataset}), strictly split into historical (PBD, $\leq$2024) and leakage-free (LFD, 2025) datasets. The framework is designed to evaluate diverse \glspl{LLM} (both vanilla and fine-tuned; \S\ref{section:models_methodology}) across three tasks (\S\ref{section:tasks}): Non-targeted Detection (Task~1), Targeted Detection (Task~2), and CWE Classification (Task~3), with performance assessed via a rigorous suite of metrics (\S\ref{section:metrics}) covering detection accuracy, pair-wise consistency, and classification ranking.}
    \label{figure:framework}
    \vspace{-6pt}
\end{figure*}

\section{Methodological Framework}
\label{section:approach}
CWE-Trace (Figure~\ref{figure:framework}) assesses security reasoning in vanilla and fine-tuned \glspl{LLM}. We extract high-fidelity Linux kernel vulnerabilities and split them into temporal splits of historical (PBD) and leakage-free (LFD) datasets, and evaluate models zero-shot on non-targeted detection, targeted detection, and hierarchical CWE classification using standard accuracy, coverage, \gls{DFI}, and \gls{HDD}. DFI quantifies directional decision bias, while HDD measures reasoning depth through shortest-path distance and error direction in the CWE taxonomy. Although this study focuses on C/Linux, the framework's extraction pipeline, temporal-split protocol, and evaluation metrics are language-agnostic: the same logic applies directly to any codebase for which commit-level CVE fixes can be retrieved and manually validated. Extension to other languages (e.g., Java, Python) is therefore a straightforward engineering task and constitutes a concrete direction for future work. The goal is to separate broad detection ability from exact semantic diagnosis under a single benchmark.

\subsection{Dataset}
\label{section:dataset}

We extract vulnerable (pre-fix) and patched (post-fix) code from Linux kernel commits and build manually labeled L1 (single-function) and L2 (multi-function) blocks. Interacting functions and non-functional elements (e.g., macros, structs, globals) are aggregated into unified instances to capture multi-file dependencies often missed by function-isolated baselines. When a fix spans multiple interacting functions, we keep those functions together in one sample; we do not label untouched files or functions as vulnerable merely because they appear in the same commit. Manual review removes non-security changes and incomplete cases. From 1,020 raw samples, two reviewers validated 628 PBD ($\leq$2024) and 206 LFD (2025) samples, excluding (1) false positives (documentation, tests, or cosmetic changes); (2) contextual gaps (missing dependencies or hardware-specific code that could not be reconstructed reliably); and (3) singleton CWEs, which provide no within-class evaluation signal and cannot support reliable CWE classification metrics, representing 2.8\% of raw samples (Table~\ref{table:dataset-stats}).

The dataset spans 74 \glspl{CWE} in the CWE-1000 hierarchy~\cite{cwe1000}. CWE-664 dominates both splits (43.3\% PBD, 60.2\% LFD); PBD also includes 12.1\% deprecated CWEs mapped to CWE-OTHER, representing historical classifications now marked as \enquote{PROHIBITED} by MITRE, while LFD concentrates on recent kernel issues such as CWE-416, 476, and 125. This composition lets us probe both taxonomic generalization on historical data and robustness on emerging vulnerabilities, rather than evaluating only one era of kernel bugs.

\subsection{Tasks and Analysis Protocol}
\label{section:tasks}
We evaluate models on three tasks that vary in guidance and answer granularity, separating broad triage behavior from exact diagnosis of weaknesses, and comparing open-ended detection with hypothesis-driven verification. All tasks use zero-shot prompting to model realistic triage conditions and ensure performance differences reflect pre-training or fine-tuning rather than prompt engineering. Structured prompting strategies (CoT, few-shot, retrieval-augmented context) may mitigate the backbone directional priors we identify, and evaluating them is a concrete direction for future work.
\subsubsection{Task 1 - Non-Targeted Vulnerability Detection}

Models receive a code block $c_i$ (vulnerable or patched) and predict $\hat{y}_i \in \{0, 1, -1\}$, where $0$~=~non-vulnerable, $1$~=~vulnerable, and $-1$~=~abstention, simulating triage when the presence and type of vulnerability are unknown:

\begin{itemize}[itemsep=2pt]
    \item \textbf{Vulnerability Detection (VD):} Correctly identifying vulnerable code samples (True Positives).
    \item \textbf{Patch Verification (PV):} Correctly identifying patched/safe code samples (True Negatives).
\end{itemize}

Abstentions reflect uncertainty and map to manual-review escalation in realistic security workflows.

\subsubsection{Task 2 - Targeted Vulnerability Detection}

Models receive a code block $c_i$ and a vulnerability hint $h_i$ (e.g., ``Check for CWE-416 Use After Free''). We evaluate \textbf{Targeted Vulnerability Detection (TVD)} and \textbf{Targeted Patch Verification (TPV)} to test whether explicit hypotheses help or whether the underlying decision policy remains unchanged. This distinguishes broad detection failures from failures of hypothesis verification.

\subsubsection{Task 3 - CWE Classification (CWE-C)}

Given vulnerable code $c_i$ with ground-truth \gls{CWE} label $y_i$, models produce a ranked list of candidate weakness types $\{\hat{y}_{i1}, \dots, \hat{y}_{ik}\}$ from the CWE-1000 taxonomy~\cite{cwe1000}. We evaluate predictions across pillars, classes, bases, and variants using \gls{HDD}, which measures shortest-path distance and error direction in the taxonomy, distinguishing shallow category-level understanding from precise fine-grained comprehension. This separates broad family recognition from exact root-cause identification and mitigates severe class imbalance in fine-grained labels. It also makes lateral, shallow, and deep errors explicit rather than collapsing them into a single ``wrong'' category.

\subsection{Models Under Analysis}
\label{section:models_methodology}
We evaluate eight off-the-shelf models in a zero-shot setting, spanning \textit{Code-Specialized} and \textit{General-Purpose} families (Table~\ref{table:llm-specs}). We also fine-tune three backbones, Qwen3-4B, DeepSeek-R1-32B, and Llama3.1-8B, on PrimeVul, LineVul, MegaVul, Devign, and VDISC, yielding 15 supervised baselines. These backbones span diverse parameter scales (4B, 8B, 32B), architectural families, and pre-training distributions, while remaining tractable for LoRA adaptation; full fine-tuning and alternative PEFT methods are directions for future work. Evaluating both PBD and LFD allows us to compare historical and post-cutoff performance under the same context-aware extraction while keeping the inference protocol fixed. This is important because the context-aware PBD presentation differs substantially from the isolated functions used in standard fine-tuning corpora, so the comparison tests not only label exposure but also whether models can transfer what they learned to a more realistic vulnerability presentation.

\begin{table*}[!t]
    \centering
    \begin{minipage}{0.28\textwidth}
        \centering
        \scriptsize
        \caption{CWE-Trace Dataset Statistics.}
        \label{table:dataset-stats}
        \setlength{\tabcolsep}{4pt}
        \renewcommand{\arraystretch}{1.1}
        \resizebox{\linewidth}{!}{%
            \begin{tabular}{@{}lrr@{}}
                \toprule
                \textbf{Metric}                 & \textbf{PBD}  & \textbf{LFD} \\
                \midrule
                Total Samples                   & 628           & 206          \\
                Unique CWEs                     & 74            & 19           \\
                \cmidrule(lr){1-3}
                \textit{Averages (per sample):} &               &              \\
                Files / Functions               & 1.5 / 2.0     & 1.1 / 1.3    \\
                Vuln / Patched Lines            & 109.4 / 114.1 & 60.9 / 63.6  \\
                Lines Added / Deleted           & 17.6 / 9.0    & 7.0 / 3.5    \\
                \bottomrule
            \end{tabular}%
        }
    \end{minipage}
    \hfill
    \begin{minipage}{0.34\textwidth}
        \centering
        \caption{\gls{CWE} Pillar Distribution in CWE-Trace.}
        \label{table:cwe-pillars-combined}
        \scriptsize
        \resizebox{\linewidth}{!}{%
            \begin{tabular}{@{}llcc@{}}
                \toprule
                \textbf{CWE-ID}    & \textbf{Pillar Name}                   & \multicolumn{2}{c}{\textbf{Coverage}}                \\
                \cmidrule(lr){3-4} &                                        & \textbf{PBD}                          & \textbf{LFD} \\
                \midrule
                CWE-664            & Improper Control of Resource Lifetime  & 136 (43.3\%)                          & 62 (60.2\%)  \\
                CWE-OTHER          & Other/Unmapped CWEs                    & 38 (12.1\%)                           & 0 (0.0\%)    \\
                CWE-682            & Incorrect Calculation                  & 32 (10.2\%)                           & 7 (6.8\%)    \\
                CWE-284            & Improper Access Control                & 28 (8.9\%)                            & 0 (0.0\%)    \\
                CWE-691            & Insufficient Control Flow Management   & 27 (8.6\%)                            & 11 (10.7\%)  \\
                CWE-707            & Improper Neutralization                & 17 (5.4\%)                            & 9 (8.7\%)    \\
                CWE-693            & Protection Mechanism Failure           & 16 (5.1\%)                            & 0 (0.0\%)    \\
                CWE-710            & Improper Adherence to Coding Standards & 11 (3.5\%)                            & 14 (13.6\%)  \\
                CWE-703            & Improper Check of Except. Conditions   & 9 (2.9\%)                             & 0 (0.0\%)    \\
                \bottomrule
            \end{tabular}%
        }
    \end{minipage}
    \hfill
    \begin{minipage}{0.34\textwidth}
        \centering
        \caption{Comparison of Fine-Tuning Datasets. \textbf{Task} denotes the supervision signal: Binary (B), Targeted detection with CWE hint (T), CWE Ranking (R).}
        \label{tab:ft-datasets}
        \scriptsize
        \setlength{\tabcolsep}{3.5pt}
        \renewcommand{\arraystretch}{1.0}
        \resizebox{\linewidth}{!}{%
            \begin{tabular}{@{}lrrrrrc@{}}
                \toprule
                \textbf{Dataset}                      & \textbf{Size} & \textbf{\# Vuln} & \textbf{\# Non-Vuln} & \textbf{CWEs} & \textbf{Train Tokens} & \textbf{Task} \\
                \midrule
                Devign~\cite{zhou2019devign}          & 27k           & 12.4k            & 14.9k                & --            & 13M                   & B             \\
                LineVul~\cite{fu2022linevul}          & 189k          & 10.9k            & 177.7k               & 91            & 191M                  & B+T+R         \\
                VDISC~\cite{russell2018automated}     & 1.29m         & 83k              & 1.21m                & 5             & 161M                  & B+T+R         \\
                PrimeVul~\cite{ding2024vulnerability} & 235k          & 7.0k             & 228.8k               & 140           & 225.2M                & B+T+R         \\
                MegaVul~\cite{ni2024megavul}          & 339k          & 17.4k            & 322.2k               & 169           & 123M                  & B+T+R         \\
                \bottomrule
            \end{tabular}%
        }
    \end{minipage}
    \vspace{-12pt}
\end{table*}

\subsection{Metrics}
\label{section:metrics}

\subsubsection{Software Vulnerability Detection}
We evaluate detection with True Positives (TP), False Positives (FP), True Negatives (TN), and False Negatives (FN), allowing abstentions ($\hat{y}_i = -1$) to represent unresolved cases:
\begin{itemize}[itemsep=2pt]
    \item \textbf{Coverage ($\text{Cov}$)}: The proportion of non-abstained predictions; low coverage indicates uncertainty:
          \begin{equation}
              \text{Cov} = \frac{N - A}{N} \times 100\%
          \end{equation}

    \item \textbf{Standard Accuracy ($\text{Acc}_{\text{std}}$)}: Accuracy treating abstentions as errors for direct cross-model comparison:
          \begin{equation}
              \text{Acc}_{\text{std}} = \frac{\text{TP} + \text{TN}}{N}
          \end{equation}

\end{itemize}

\subsubsection{CWE Classification}
We assess ranking quality and semantic proximity with the following metrics:
\begin{itemize}[itemsep=2pt]
    \item \textbf{Top-$k$ Accuracy}~\cite{russakovsky2015imagenet}: Prediction is correct if the true label $y_i$ appears in the top-$k$ ranked predictions $\{\hat{y}_{i1}, \dots, \hat{y}_{ik}\}$:
          \begin{equation}
              \mathrm{Acc}@k = \frac{1}{N} \sum_{i=1}^{N} \mathbb{I}\left[y_i \in \{\hat{y}_{i1}, \dots, \hat{y}_{ik}\}\right]
          \end{equation}
          where $\mathbb{I}[\cdot]$ is the indicator function. We report Top-1 through Top-5 to evaluate both exact classification and ranking quality, both of which are relevant to security triage.

    \item \textbf{\gls{MRR}}~\cite{craswell2009mean}: Average reciprocal rank of the correct label:
          \begin{equation}
              \text{MRR} = \frac{1}{N}\sum_{i=1}^{N}\frac{1}{\text{rank}_i}
          \end{equation}
          where $\text{rank}_i$ is the position of $y_i$ in the ranked predictions. Higher MRR indicates better prioritization of the correct vulnerability type, even when the top prediction is wrong.

    \item \textbf{Root-Taxonomy Accuracy (RootMicro@1 / RootMacro@1)}: We map each prediction to its \gls{CWE}-1000 root pillar and compute:
          \begin{align}
              \text{RootMicro@1} & = \frac{1}{N}\sum_{i=1}^{N}\mathbb{I}\!\left[\hat{r}_i = r_i\right]                                                      \\[2pt]
              \text{RootMacro@1} & = \frac{1}{|\mathcal{C}|}\sum_{c \in \mathcal{C}} \frac{1}{N_c}\sum_{i:\, r_i = c}\mathbb{I}\!\left[\hat{r}_i = c\right]
          \end{align}
          where $r_i$ is the ground-truth root pillar, $\hat{r}_i$ is the predicted root, $\mathcal{C}$ is the set of root classes, and $N_c = |\{i : r_i = c\}|$. RootMicro@1 is instance-weighted and dominated by frequent families; RootMacro@1 weights roots equally, rewarding performance on rarer pillars.

    \item \textbf{\gls{HDD}}: A taxonomy-aware metric defined by two quantities. For a prediction $p$ and ground-truth CWE $g$:
          \begin{equation}
              \text{Distance}(p, g) = \text{up}(p, \text{LCA}) + \text{down}(\text{LCA}, g)
          \end{equation}
          where $\text{LCA}$ is the lowest common ancestor of $p$ and $g$ in the CWE hierarchy, and $\text{up}/\text{down}$ count taxonomy edges to and from the LCA. Thus the unit of HDD distance is one edge in the CWE-1000 graph.
          \begin{equation}
              \text{DirectionGap}(p, g) = \text{depth}(p) - \text{depth}(g)
          \end{equation}
          A negative gap indicates a shallower prediction; a positive gap indicates a deeper one. Each prediction is classified as \textit{exact} ($p=g$), \textit{shallow} (ancestor of $g$), \textit{deep} (descendant of $g$), \textit{lateral} (same-depth or sibling-branch error under a shared ancestor), or \textit{invalid} (not mapped to CWE-1000). We report mean distance ($\overline{d}$), mean specificity gap, coverage, and the full direction profile so that partial taxonomy understanding remains distinguishable from unrelated error.
\end{itemize}

\subsubsection{Diagnostic Metrics}
Two additional diagnostic indices characterize decision biases beyond aggregate accuracy:
\begin{itemize}[itemsep=2pt]
    \item \textbf{Directional Failure Index (DFI)}: A signed metric for decision asymmetry ($\delta$):
          \begin{equation}
              \delta = \text{Acc}_{\text{vuln}} - \text{Acc}_{\text{safe}} = \frac{\text{TP}}{\text{TP}+\text{FN}} - \frac{\text{TN}}{\text{TN}+\text{FP}}
          \end{equation}
           In our four-task evaluation, this is computed as:
          \begin{equation}
              \delta = \frac{\text{VD} + \text{TVD}}{2} - \frac{\text{PV} + \text{TPV}}{2}
          \end{equation}
          A positive $\delta$ indicates a \textit{paranoid} bias; a negative $\delta$ indicates a \textit{skeptical} bias. A balanced model approaches $\delta \approx 0$, so large magnitudes reveal directional collapse even when average accuracy is near chance.

    \item \textbf{Contextual Gain Ratio (CGR)}: Measures the performance uplift ($\rho$) from vulnerability hints:
          \begin{align}
              \rho                         & = \text{Acc}_{\text{targeted}} - \text{Acc}_{\text{base}}, \\
              \text{Acc}_{\text{base}}     & = \frac{\text{VD}+\text{PV}}{2},                            \\
              \text{Acc}_{\text{targeted}} & = \frac{\text{TVD}+\text{TPV}}{2}.
          \end{align}
          A positive $\rho$ indicates that the model benefits from the CWE hint, narrowing the gap between open-ended detection and hypothesis verification.
\end{itemize}

\section{Experimental Setup}
\label{section:setup}

We evaluate the eight vanilla models listed in Table~\ref{table:llm-specs} using their official checkpoints. All models are run in \texttt{bfloat16} precision. To ensure reproducibility against post-release updates, we pin open-weight models to the specific HuggingFace commit SHAs detailed in Table~\ref{table:llm-specs} and deploy frozen local checkpoints. All inference runs use temperature~$= 0$; the directional biases we report (DFI ranging from $-85.5$ to $+94.8$~pp) substantially exceed sampling variance at moderate temperatures, confirming that results are representative of each model's dominant decision policy.

For the supervised baselines, we fine-tune three backbones, \texttt{Qwen3-4B}, \texttt{DeepSeek-R1-32B}, and \texttt{Llama3.1-8B}, on five SOTA vulnerability datasets, \texttt{PrimeVul}, \texttt{LineVul}, \texttt{MegaVul}, \texttt{Devign}, and \texttt{VDISC}, using a parameter-efficient LoRA strategy~\cite{hu2022lora}. This 3$\times$5 design yields 15 fine-tuned variants. All variants use rank $r{=}16$, $\alpha{=}32$, dropout $= 0$, applied to all attention and feed-forward projections (\texttt{q}, \texttt{k}, \texttt{v}, \texttt{o}, \texttt{gate}, \texttt{up}, \texttt{down\_proj}), with learning rate $2{\times}10^{-4}$, cosine decay, $5\%$ warmup, weight decay $0.01$, gradient clipping $1.0$, and batch size 64. These are standard LoRA defaults following~\cite{hu2022lora}; sensitivity to rank and $\alpha$ is a direction for future work. Training converges within one epoch.

Training corpora are skewed toward safe samples (PrimeVul 4.6\%, MegaVul 5.1\%, LineVul/VDISC ${\sim}6.5\%$, Devign 45.5\% vulnerable). We apply sample-level oversampling of vulnerable examples (15--20$\times$ for PrimeVul/MegaVul/VDISC, $2\times$ for Devign, disabled for LineVul, whose multitask quota already balances classes). We use oversampling rather than scalar loss reweighting because prior work shows reweighting alone fails to resolve semantic conflict in noisy vulnerability datasets~\cite{ding2024vulnerability}.

\footnotetext[1]{The Qwen3 technical report~\cite{yang2025qwen3} does not specify an exact training data cutoff. The range ``late 2024--early 2025'' is inferred from the model release date (April 2025) and the pre-training data description in the report, which is the standard basis used by the community for this model.}

To isolate dataset quality, we apply the same three-task objective to all datasets with CWE labels: \textit{binary detection} (VD/PV), \textit{targeted detection with CWE hint} (TVD/TPV), and \textit{CWE ranking}. Devign, which has no CWE labels, is trained with binary detection only. This keeps the task fixed across datasets, so differences reflect the supervision signal rather than different objectives.

Known overlap between PBD and the fine-tuning datasets serves as a control variable when comparing historical and leakage-free performance. High performance on overlapping samples combined with a drop in LFD would indicate memorization rather than generalization. The absence of that pattern is therefore also informative, because it helps distinguish nominal overlap from function-level exposure.

We report Standard Accuracy, treating abstentions as incorrect. Coverage and DFI are reported alongside each other to provide a faithful view of practical triage behavior.

\begin{table}[!t]
    \centering
    \caption{LLMs with training cutoffs and contamination controls.}
    \label{table:llm-specs}
    \vspace{-6pt}
    \scriptsize
    \setlength{\tabcolsep}{3pt}
    \renewcommand{\arraystretch}{1.15}
    \resizebox{\columnwidth}{!}{%
        \begin{tabular}{@{}llrrlllc@{}}
            \toprule
            \textbf{Domain} & \textbf{Model}                           & \textbf{Size} & \textbf{Ctx} & \textbf{Cutoff}                  & \textbf{HF Commit Date} & \textbf{HF SHA}   & \textbf{Tier} \\
            \midrule
            \multirow{3}{*}{\textbf{Code-Spec.}}
                            & CodeLlama~\cite{roziere2023code}         & 7B            & 16K          & $\sim$Sep 2022                   & 2024-04-12              & \texttt{6c284d14} & 1             \\
                            & StarCoder2~\cite{lozhkov2024starcoder}   & 7B            & 16K          & Sep 2023                         & 2024-06-11              & \texttt{bb9afde7} & 1             \\
                            & Qwen3-Coder~\cite{yang2025qwen3}         & 30B           & 262K         & Dec 2024                         & 2025-07-31              & \texttt{d4add6db} & 2             \\
            \cmidrule{1-8}
            \multirow{5}{*}{\textbf{General}}
                            & Mistral v0.3~\cite{jiang2023mistral}     & 7B            & 32K          & $\sim$2023                       & 2024-08-21              & \texttt{e0bc86c2} & 1             \\
                            & Llama3.1~\cite{touvron2023llama}           & 8B            & 131K         & Dec 2023                         & 2024-09-25              & \texttt{0e9e39f2} & 1             \\
                            & DeepSeek-R1~\cite{guo2025deepseek}       & 7B            & 131K         & Dec 2023                         & 2025-01-20              & \texttt{4e5485ed} & 2             \\
                            & Qwen3~\cite{yang2025qwen3}               & 4B            & 262K         & $\sim$Early 2025\footnotemark[1] & 2025-04-28              & \texttt{9e1b55c7} & 2             \\
                            & GPT-4.1-mini~\cite{openai2025gpt4_1mini} & --            & 1M           & Jun 2024                         & 2025-04-14              & (API)             & 2             \\
            \cmidrule{1-8}
            \multirow{3}{*}{\textbf{FT Base}}
                            & Qwen3~\cite{yang2025qwen3}               & 4B            & 262K         & $\sim$Early 2025\footnotemark[1] & 2025-04-28              & \texttt{9e1b55c7} & 2             \\
                            & DeepSeek-R1~\cite{guo2025deepseek}       & 32B           & 131K         & Dec 2023                         & 2025-01-20              & \texttt{d66bcfc2} & 2             \\
                            & Llama3.1~\cite{touvron2023llama}           & 8B            & 131K         & Dec 2023                         & 2024-09-25              & \texttt{0e9e39f2} & 1             \\
            \bottomrule
        \end{tabular}%
    }
    \vspace{-12pt}
\end{table}
\textbf{Research Questions.} We consider three research questions:

\begin{itemize}
    \item \textbf{RQ1 (Vulnerability Detection Reliability):} How reliably do vanilla and fine-tuned models detect vulnerabilities and verify patches, and what roles do directional bias, label-space transfer, and contamination play?
    \item \textbf{RQ2 (CWE-1000 Classification Transfer):} Does fine-tuning improve coarse-grained \gls{CWE}-1000 / root-taxonomy classification, and do these gains transfer to leakage-free future vulnerabilities?
    \item \textbf{RQ3 (Semantic Depth of Security Understanding):} Do models move beyond broad vulnerability families to exact, hierarchy-aware \gls{CWE} reasoning, or do they remain limited to shallow taxonomic proximity?
\end{itemize}

\section{RQ1: Vulnerability Detection Performance and Directional Failure}
\label{section:rq1}

We begin by evaluating detection performance of vanilla and fine-tuned models on PBD and LFD. Table~\ref{tab:rq1-vd-overall} summarizes results for non-targeted (VD/PV) and targeted (TVD/TPV) detection. Figure~\ref{fig:rq1_mechanisms} then isolates two mechanism questions that Table~\ref{tab:rq1-vd-overall} cannot answer: whether fine-tuning helps on the \gls{CWE} label space it saw during training, and whether it benefits from contaminated ``seen-CVE'' samples under our PBD extraction.

\begin{table*}[!b]
    \vspace{-12pt}
    \centering
    \caption{Vulnerability detection performance (standard accuracy, \%) on PBD and LFD. DFI measures decision bias (positive: paranoid/high-FP; negative: skeptical/high-FN). CGR denotes the effect of CWE context; Gen $=$ LFD $-$ PBD. Overall is the mean of PBD Avg and LFD Avg, and Rank is based on Overall ($1$ = best). Per-section \textit{MEAN} rows summarize each block; in the Rank column, they report the average rank.}
    \label{tab:rq1-vd-overall}
    \vspace{-6pt}
    \scriptsize
    \setlength{\tabcolsep}{3.5pt}
    \renewcommand{\arraystretch}{1.1}
    \begin{tabular}{l|rrrrrr|rrrrrr|rrrrrr}
        \toprule
                       & \multicolumn{6}{c|}{\textbf{PBD (Historical $\leq$ 2024)}} & \multicolumn{6}{c|}{\textbf{LFD (Leakage-Free 2025)}} & \multicolumn{6}{c}{\textbf{Summary}}                                                                                                                                                                                                                                                     \\
        \cmidrule(lr){2-7}\cmidrule(lr){8-13}\cmidrule(lr){14-19}
        \textbf{Model} & \textbf{VD}                                                & \textbf{PV}                                           & \textbf{TVD}                         & \textbf{TPV} & \textbf{Avg} & \textbf{DFI} & \textbf{VD} & \textbf{PV} & \textbf{TVD} & \textbf{TPV} & \textbf{Avg} & \textbf{DFI} & \textbf{CGR} & \textbf{Gen} & \textbf{Cov$_{PBD}$} & \textbf{Cov$_{LFD}$} & \textbf{Overall} & \textbf{Rank} \\
        \midrule
        \multicolumn{19}{c}{\textit{Vanilla Models}}                                                                                                                                                                                                                                                                                                                                                                                   \\
        \midrule
        StarCoder2     & 16.2                                                       & 64.7                                                  & 32.2                                 & 50.0         & 40.8         & -33.1        & 10.7        & 68.0        & 30.1         & 56.3         & 41.3         & -41.7        & +2.3         & +0.5         & 85.9                 & 78.2                 & 41.1             & 18            \\
        Qwen3-4B       & 1.6                                                        & 36.6                                                  & 0.0                                  & 18.8         & 14.3         & -26.9        & 4.9         & 29.1        & 0.0          & 3.9          & 9.5          & -14.1        & -9.7         & -4.8         & 36.9                 & 32.0                 & 11.9             & 22            \\
        Mistral        & 88.2                                                       & 10.5                                                  & 20.1                                 & 73.9         & 48.2         & +11.9        & 80.6        & 21.4        & 14.6         & 82.5         & 49.8         & -4.4         & -2.4         & +1.6         & 100.0                & 100.0                & 49.0             & 8             \\
        DeepSeek-R1    & 96.1                                                       & 4.5                                                   & 98.7                                 & 0.7          & 50.0         & +94.8        & 91.3        & 8.7         & 99.0         & 0.0          & 49.8         & +90.8        & -0.6         & -0.2         & 100.0                & 100.0                & 49.9             & 6             \\
        GPT-4.1-mini   & 5.7                                                        & 93.6                                                  & 8.0                                  & 91.1         & 49.6         & -85.5        & 4.8         & 96.1        & 8.7          & 96.1         & 51.5         & -89.3        & +0.9         & +1.9         & 98.3                 & 97.3                 & 50.6             & 3             \\
        Llama3.1       & 31.2                                                       & 68.2                                                  & 26.8                                 & 72.6         & 49.7         & -41.4        & 35.9        & 76.7        & 13.6         & 79.6         & 51.5         & -53.4        & -4.9         & +1.8         & 100.0                & 100.0                & 50.6             & 3             \\
        Qwen3-Coder    & 84.4                                                       & 16.6                                                  & 61.1                                 & 43.0         & 51.3         & +43.0        & 80.6        & 19.4        & 47.6         & 54.4         & 50.5         & +27.2        & +1.3         & -0.8         & 99.1                 & 99.5                 & 50.9             & 2             \\
        CodeLlama      & 75.9                                                       & 21.2                                                  & 92.3                                 & 11.2         & 50.2         & +67.8        & 70.9        & 31.1        & 92.2         & 21.4         & 53.9         & +55.3        & +4.5         & +3.7         & 99.6                 & 100.0                & 52.1             & 1             \\
        \cmidrule(lr){1-19}
        \textbf{MEAN}  & 49.9                                                       & 39.5                                                  & 42.4                                 & 45.2         & 44.3         & +3.8         & 47.5        & 43.8        & 38.2         & 49.3         & 44.7         & -3.7         & -0.5         & +0.4         & 92.0                 & 89.9                 & 44.5             & --            \\
        \midrule
        \multicolumn{19}{c}{\textit{Fine-Tuned Models (Base: Qwen3-4B)}}                                                                                                                                                                                                                                                                                                                                                               \\
        \midrule
        MegaVul-FT     & 22.6                                                       & 51.3                                                  & 35.7                                 & 57.3         & 41.7         & -25.2        & 24.3        & 42.7        & 36.9         & 50.5         & 38.6         & -16.0        & +9.5         & -3.1         & 80.3                 & 73.5                 & 40.2             & 19            \\
        PrimeVul-FT    & 12.7                                                       & 80.6                                                  & 10.8                                 & 79.6         & 45.9         & -68.3        & 16.5        & 72.8        & 9.7          & 91.3         & 47.6         & -68.9        & -1.5         & +1.7         & 91.6                 & 95.4                 & 46.8             & 14            \\
        VDISC-FT       & 45.2                                                       & 46.5                                                  & 8.9                                  & 85.0         & 46.4         & -38.7        & 39.8        & 57.3        & 5.8          & 88.3         & 47.8         & -50.0        & +1.1         & +1.4         & 91.8                 & 91.7                 & 47.1             & 13            \\
        Devign-FT      & 1.0                                                        & 91.7                                                  & 1.3                                  & 92.4         & 46.6         & -90.9        & 1.0         & 95.1        & 2.9          & 95.1         & 48.5         & -93.2        & +0.5         & +1.9         & 93.8                 & 95.6                 & 47.6             & 11            \\
        LineVul-FT     & 22.3                                                       & 75.2                                                  & 24.2                                 & 77.1         & 49.7         & -52.9        & 25.2        & 74.8        & 30.1         & 73.8         & 51.0         & -46.6        & +1.9         & +1.3         & 96.3                 & 99.0                 & 50.4             & 5             \\
        \cmidrule(lr){1-19}
        \textbf{MEAN}  & 15.3                                                       & 72.7                                                  & 20.4                                 & 75.6         & 46.0         & -56.3        & 16.8        & 70.9        & 23.3         & 73.1         & 46.0         & -51.9        & +4.2         & +0.0         & 90.1                 & 89.4                 & 46.0             & 12.4          \\
        \midrule
        \multicolumn{19}{c}{\textit{Fine-Tuned Models (Base: Llama3.1 8B)}}                                                                                                                                                                                                                                                                                                                                                            \\
        \midrule
        Devign-FT      & 0.0                                                        & 4.1                                                   & 0.0                                  & 3.2          & 1.8          & -3.7         & 0.0         & 1.9         & 0.0          & 1.0          & 0.7          & -1.5         & -0.4         & -1.1         & 4.1                  & 1.5                  & 1.3              & 23            \\
        LineVul-FT     & 0.3                                                        & 89.8                                                  & 2.9                                  & 89.5         & 45.6         & -88.1        & 1.0         & 81.6        & 0.0          & 94.2         & 44.2         & -87.4        & +1.1         & -1.4         & 92.3                 & 90.5                 & 44.9             & 15            \\
        MegaVul-FT     & 22.0                                                       & 77.7                                                  & 21.7                                 & 72.3         & 48.4         & -53.2        & 16.5        & 82.5        & 15.5         & 70.9         & 46.4         & -60.7        & -2.9         & -2.0         & 94.8                 & 94.9                 & 47.4             & 12            \\
        PrimeVul-FT    & 1.0                                                        & 97.8                                                  & 2.5                                  & 90.4         & 47.9         & -92.4        & 0.0         & 100.0       & 3.9          & 93.2         & 49.3         & -94.7        & -2.9         & +1.4         & 95.5                 & 96.1                 & 48.6             & 9             \\
        VDISC-FT       & 90.4                                                       & 7.3                                                   & 94.9                                 & 5.4          & 49.5         & +86.3        & 97.1        & 1.9         & 98.1         & 1.9          & 49.8         & +95.6        & +1.3         & +0.3         & 99.0                 & 100.0                & 49.7             & 7             \\
        \cmidrule(lr){1-19}
        \textbf{MEAN}  & 22.7                                                       & 55.4                                                  & 24.4                                 & 52.2         & 38.7         & -30.2        & 22.9        & 53.6        & 23.5         & 52.2         & 38.1         & -29.7        & -0.6         & -0.6         & 77.2                 & 76.6                 & 38.4             & 13.2          \\
        \midrule
        \multicolumn{19}{c}{\textit{Fine-Tuned Models (Base: DeepSeek-R1-32B)}}                                                                                                                                                                                                                                                                                                                                                        \\
        \midrule
        MegaVul-FT     & 2.5                                                        & 59.2                                                  & 0.3                                  & 21.3         & 20.9         & -38.9        & 0.0         & 88.3        & 0.0          & 16.5         & 26.2         & -52.4        & -28.0        & +5.4         & 40.9                 & 51.5                 & 23.6             & 22            \\
        VDISC-FT       & 48.4                                                       & 40.4                                                  & 1.6                                  & 72.6         & 40.8         & -31.5        & 11.7        & 45.6        & 0.0          & 41.7         & 24.8         & -37.9        & -7.3         & -16.0        & 81.9                 & 49.3                 & 32.8             & 21            \\
        Devign-FT      & 42.0                                                       & 51.0                                                  & 7.3                                  & 85.4         & 46.4         & -43.5        & 29.1        & 59.2        & 0.0          & 0.0          & 22.1         & -15.0        & -22.2        & -24.3        & 95.8                 & 49.8                 & 34.3             & 20            \\
        PrimeVul-FT    & 1.3                                                        & 83.4                                                  & 0.0                                  & 90.4         & 43.8         & -86.3        & 0.0         & 92.2        & 0.0          & 82.5         & 43.7         & -87.4        & +2.9         & -0.1         & 87.0                 & 87.4                 & 43.8             & 16            \\
        LineVul-FT     & 0.3                                                        & 98.4                                                  & 0.0                                  & 99.0         & 49.4         & -98.6        & 0.0         & 93.2        & 0.0          & 97.1         & 47.6         & -95.1        & +0.2         & -1.8         & 97.5                 & 90.0                 & 48.5             & 10            \\
        \cmidrule(lr){1-19}
        \textbf{MEAN}  & 18.9                                                       & 66.5                                                  & 1.8                                  & 73.8         & 40.3         & -59.7        & 8.2         & 75.7        & 0.0          & 47.6         & 32.9         & -57.6        & -11.5        & -7.4         & 80.6                 & 65.6                 & 36.6             & 17.8          \\
        \bottomrule
    \end{tabular}
\end{table*}

Before examining \glspl{LLM}, Table~\ref{tab:non-llm-baselines} situates them against non-\gls{LLM} reference points. Semgrep has near-zero recall ($0.3\%$ on PBD, $0.0\%$ on LFD), confirming that pattern matching cannot recover the semantic root causes in CWE-Trace. DeepWukong predicts all samples as safe. On the $\leq$512-token subset, GraphCodeBERT reaches $61.5\%$ F1 on PBD and $62.5\%$ on LFD, but this remains far below what would be needed on the full long-context benchmark. These baselines reinforce that the binding constraint is not only the model family but also the ability to ingest the context that realistic vulnerabilities require.

\begin{table}[!bp]
    \vspace{-12pt}
    \centering
    \caption{Baselines on CWE-Trace. Models evaluated on the $\leq$512-token subset only (30.6\% of PBD and 46.6\% of LFD), while Semgrep and DeepWukong utilize the full evaluation splits.}
    \label{tab:non-llm-baselines}
    \footnotesize
    \setlength{\tabcolsep}{5pt}
    \renewcommand{\arraystretch}{1.15}
    \resizebox{\columnwidth}{!}{%
        \begin{tabular}{@{}lrrrrrrrr@{}}
            \toprule
                                                      & \multicolumn{4}{c}{\textbf{PBD}} & \multicolumn{4}{c}{\textbf{LFD}}                                                                                           \\
            \cmidrule(lr){2-5} \cmidrule(l){6-9}
            \textbf{Baseline}                         & \textbf{Acc}                     & \textbf{Rec}                     & \textbf{Prec} & \textbf{F1} & \textbf{Acc} & \textbf{Rec} & \textbf{Prec} & \textbf{F1} \\
            \midrule
            Semgrep~\cite{semgrep}                    & 49.7                             & 0.3                              & 33.3          & 0.6         & 50.2         & 0.0          & ---           & 0.0         \\
            LineVul~\cite{fu2022linevul}              & 47.4                             & 0.0                              & ---           & 0.0         & 49.0         & 4.1          & 50.0          & 7.6         \\
            ReVeal~\cite{chakraborty2021deep}         & 47.9                             & 14.0                             & 50.0          & 21.9        & 49.0         & 4.1          & 50.0          & 7.6         \\
            GraphCodeBERT~\cite{guo2020graphcodebert} & 50.5                             & 76.0                             & 51.7          & 61.5        & 50.0         & 81.6         & 50.6          & 62.5        \\
            DeepWukong~\cite{cheng2021deepwukong}     & 47.9                             & 0.0                              & ---           & 0.0         & 49.0         & 0.0          & ---           & 0.0         \\
            \bottomrule
        \end{tabular}%
    }
    \vspace{3pt}
    \vspace{-4pt}
\end{table}

\subsection{Finding 1: Base Models Perform Near-Random with Severe Directional Bias}
\label{section:rq1_bias}

Table~\ref{tab:rq1-vd-overall} shows that overall accuracy clusters at $49$--$53\%$ across most vanilla models, where random chance is $50\%$. The best vanilla model, CodeLlama, reaches only $52.1\%$ Overall ($+2.1$~pp above chance). However, near-$50\%$ averages mask extreme \emph{directional} failure rather than balanced reasoning. DeepSeek-R1 is strongly \textit{paranoid} ($\delta = +94.8$~pp on PBD, $+90.8$~pp on LFD), effectively flagging code as vulnerable and accumulating true positives at the cost of false positives. GPT-4.1-mini is the opposite: strongly \textit{skeptical} ($\delta = -85.5$~pp on PBD, $-89.3$~pp on LFD), almost always predicting ``safe'' and missing nearly every vulnerability. Qwen3-4B occupies a third failure mode: near-total abstention, with 36.9\% PBD coverage and 11.9\% Overall. These patterns persist on LFD, indicating stable backbone priors rather than contamination effects.

\subsection{Finding 2: Fine-Tuning Effect Is Determined by the Base Model's Initial Competence}
\label{section:rq1_ft_effect}

Fine-tuning does not uniformly improve detection; direction and magnitude track the backbone's starting condition. For \textit{Qwen3-4B}, performance rises by $+27.5$ to $+41.5$~pp because fine-tuning mainly overcomes abstention: coverage increases from $36.9\%$ to above $91\%$, so the apparent gain reflects producing binary verdicts rather than learning transferable vulnerability semantics. For \textit{Llama3.1}, fine-tuning is mostly neutral or harmful ($-0.2$ to $-50.7$~pp), with Devign-FT collapsing to $1.3\%$ Overall and $4.1\%$ coverage on PBD. For \textit{DeepSeek-R1}, fine-tuning consistently degrades detection ($-0.6$ to $-29.1$~pp on LFD), often amplifying its paranoid prior. The hardest backbone to improve in detection benefits most in RQ2 coarse \gls{CWE} classification, reinforcing that detection and understanding are weakly coupled.

\begin{figure}[!t]
    \centering
    \includegraphics[width=\columnwidth]{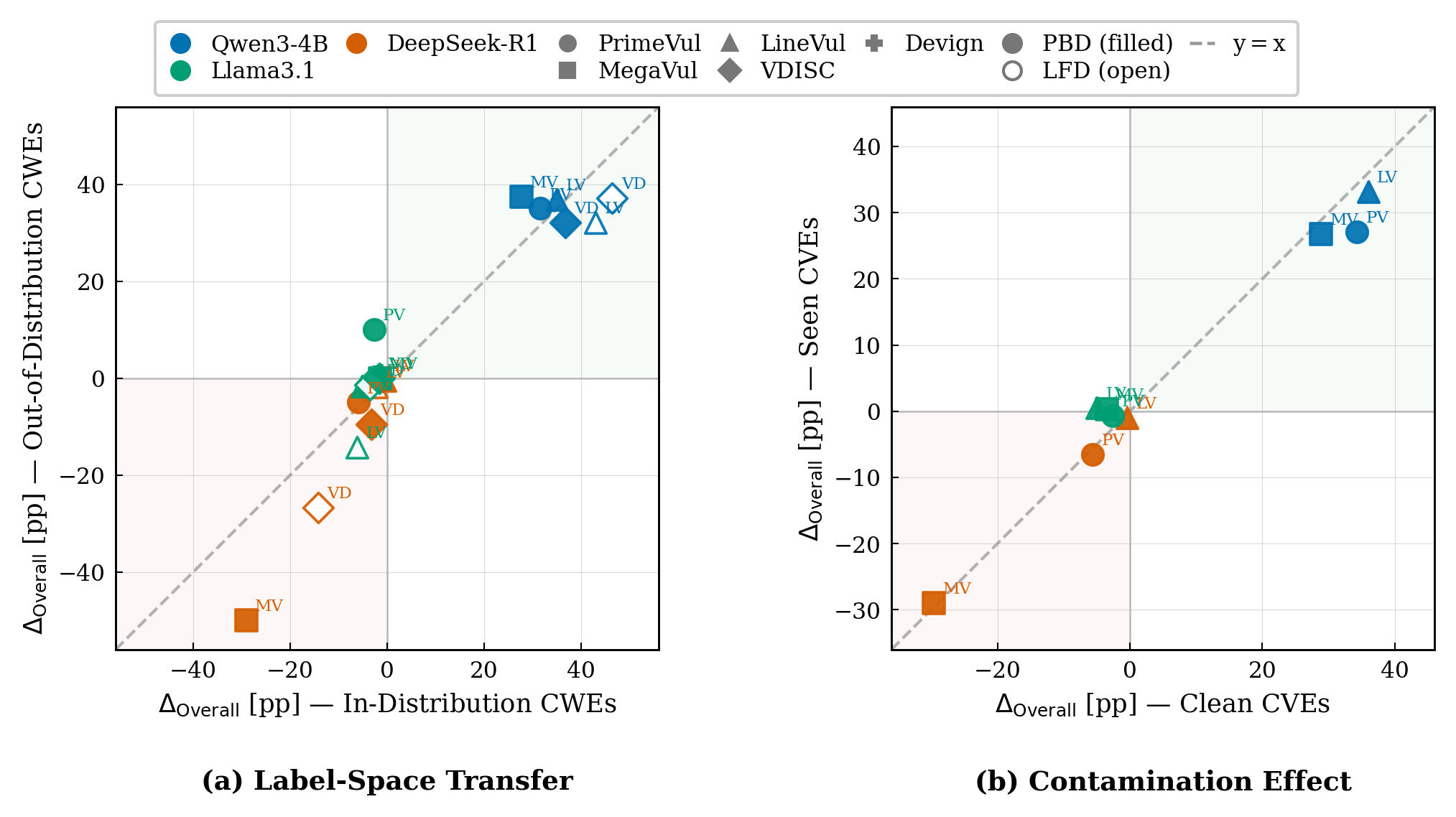}
    \vspace{-18pt}
    \caption{RQ1 mechanism analysis. Panel~(a): fine-tuning vs.\ backbone detection deltas on samples whose ground-truth \gls{CWE} falls inside vs.\ outside the fine-tuning label space. Panel~(b): deltas on contaminated (seen-CVE) vs.\ clean PBD samples. Each point is one fine-tuned model on PBD or LFD; positive values indicate improved Overall accuracy over the vanilla backbone.}
    \label{fig:rq1_mechanisms}
    \vspace{-12pt}
\end{figure}

\subsection{Finding 3: Label-Space Transfer Is Flat---Fine-Tuning Shifts Policy, Not CWE Knowledge}
\label{section:rq1_label_space}

Figure~\ref{fig:rq1_mechanisms}(a) shows little dependence on whether a sample falls inside the fine-tuning dataset's labeled \gls{CWE} space. For Qwen3-4B, gains on covered and uncovered \glspl{CWE} are nearly identical on LFD, indicating a global response-policy shift rather than CWE-specific learning. Llama3.1-8B and DeepSeek-R1-32B regress on both subsets. Fine-tuning, therefore, does not reliably help most where the training labels should matter most.

\subsection{Finding 4: Contamination Has No Detectable Effect}
\label{section:rq1_contamination}

Figure~\ref{fig:rq1_mechanisms}(b) shows no consistent advantage of contaminated seen-CVE samples over clean ones. Qwen3-4B improves slightly more on clean CVEs ($+36.0$ vs.\ $+33.2$~pp for LineVul-FT), Llama3.1-8B is near-zero on seen CVEs but negative on clean ones, and DeepSeek-R1-32B is negative on both. Contamination provides no memorization shortcut; Section \ref{sec:data_quality} explains why through function-level distribution shift and semantic-conflict label noise.

\subsection{Finding 5: Context Depth Is Necessary but Not Sufficient}
\label{sec:l1l2}

CWE-Trace distinguishes \textbf{L1} samples (single file, single function; $n_\text{PBD}=207$, $n_\text{LFD}=84$) from \textbf{L2} samples (multi-file or multi-function; $n_\text{PBD}=104$, $n_\text{LFD}=17$). Table~\ref{tab:l1l2} shows almost no split-level difference: on PBD, both groups are flat across context depth ($+2.3$~pp and $+1.6$~pp), and the LFD vanilla gap ($-2.1$~pp) is not statistically reliable at $n_{\text{LFD-L2}}=17$. Near-chance performance persists across both depths, indicating that context depth alone does not overcome dominant decision priors.


RQ1 yields a stronger conclusion than ``accuracy is low'': stable directional priors dominate the decision policy, fine-tuning mainly shifts these priors rather than learning robust, transferable behavior, and neither labeled \gls{CWE} overlap nor contaminated seen-CVE overlap provides a reliable path to improved detection.

\begin{table}[!t]
    \centering
    \scriptsize
    \setlength{\tabcolsep}{4pt}
    \renewcommand{\arraystretch}{1.05}
    \caption{Average overall accuracy (\%) by context depth (L1/L2), split, and model group.}
    \label{tab:l1l2}
    \begin{tabular}{llccc}
        \toprule
        \textbf{Group}                   & \textbf{Split} & \textbf{L1 Acc} & \textbf{L2 Acc} & $\boldsymbol{\Delta}$ \textbf{(L2$-$L1)} \\
        \midrule
        \multirow{2}{*}{Vanilla ($n$=8)} & PBD            & 44.0            & 46.2            & $+2.3$                                   \\
                                         & LFD            & 46.2            & 44.1            & $-2.1$                                   \\
        \midrule
        \multirow{2}{*}{FT ($n$=15)}     & PBD            & 41.7            & 43.3            & $+1.6$                                   \\
                                         & LFD            & 42.0            & 41.2            & $-0.8$                                   \\
        \bottomrule
    \end{tabular}
    \vspace{-12pt}
\end{table}

\section{RQ2: CWE-1000 Classification Before and After Fine-Tuning}
\label{section:rq2}

\begin{figure*}[!t]
    \centering
    \includegraphics[width=0.8\textwidth]{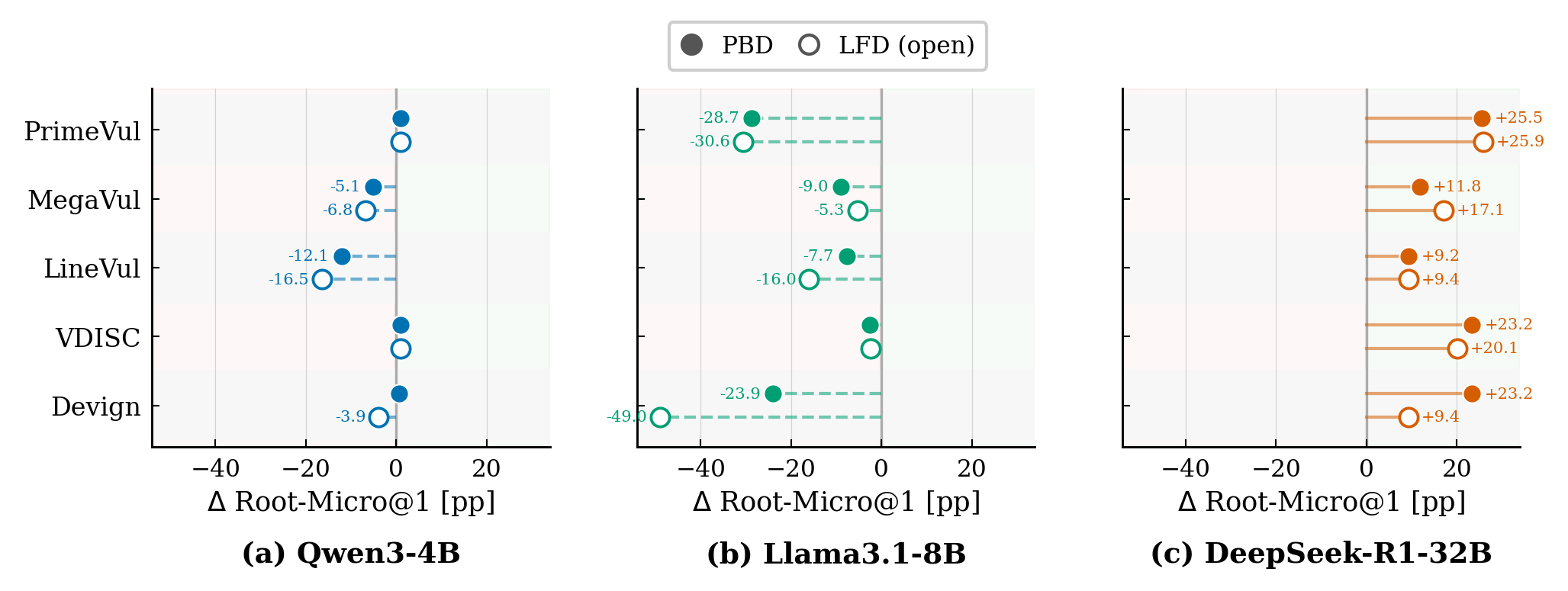}
    \vspace{-6pt}
    \caption{Fine-tuning $\Delta$ Root-Micro@1 for coarse \gls{CWE}-1000 / root-taxonomy classification (relative to vanilla backbone). Each lollipop represents the change for a single fine-tuning dataset; filled circles = PBD, open circles = LFD. Red zone: fine-tuning hurts; green zone: fine-tuning helps.}
    \label{fig:rq2_taxonomy}
    \vspace{-12pt}
\end{figure*}

RQ2 asks whether models can place vulnerabilities into the correct \emph{coarse} \gls{CWE}-1000 root family, a weaker target than exact \gls{CWE} identification, capturing broad security understanding before probing semantic depth in RQ3.

\subsection{Finding 1: Base Models Reach Moderate Coarse-Level Accuracy but With Macro Disparity}
\label{section:rq2_base}

Vanilla models perform materially better at root classification than at exact \gls{CWE} ranking. StarCoder2 leads on LFD at $61.0\%$ Micro@1 and Qwen3-4B follows at $59.2\%$, but Macro@1 remains much lower ($9.7$--$19.3\%$ on PBD), revealing strong bias toward dominant root families. GPT-4.1-mini reaches the best LFD Macro@1 at $23.8\%$, suggesting more balanced predictions even without the best Micro@1. The Macro gap shows that even at the coarse level, models still concentrate predictions on a narrow set of frequent roots rather than reasoning across the full taxonomy.

\subsection{Finding 2: Fine-Tuning Response Is Systematically Backbone-Dependent and Inverse to RQ1}
\label{section:rq2_backbone}

Figure~\ref{fig:rq2_taxonomy} shows a consistent backbone signature. \textit{DeepSeek-R1-32B} improves on every fine-tuning dataset ($+9.2$ to $+25.9$~pp on PBD; $+9.4$ to $+25.9$~pp on LFD), whereas \textit{Llama3.1-8B} degrades on every one ($-2.4$ to $-28.7$~pp on PBD; $-2.4$ to $-49.0$~pp on LFD). \textit{Qwen3-4B} is near-neutral on average but harmed by LineVul-FT. Backbone identity, not the fine-tuning recipe, determines transfer direction. This is also the inverse of RQ1: the backbone most harmed in detection can still gain most on coarse taxonomy placement.

\subsection{Finding 3: Binary-Label Fine-Tuning Causes Catastrophic Forgetting of CWE Knowledge}
\label{section:rq2_forgetting}

The clearest evidence of catastrophic forgetting is Llama3.1-8B Devign-FT, which drops $-49.0$~pp on LFD Root-Micro@1, the worst single result in the study. Because Devign has no \gls{CWE} labels, binary-only fine-tuning can erase coarse taxonomy knowledge rather than improve it.

RQ2 shows that coarse \gls{CWE}-1000 gains are possible, but only for certain backbones and at a family-level granularity; the effect is thus backbone-dependent rather than a general recipe. Fine-tuning \emph{can} improve coarse \gls{CWE}-1000 placement, but not consistently across models: the same intervention may degrade another backbone, and binary-label datasets can even erase existing taxonomy knowledge.

\section{RQ3: Semantic Depth and Hierarchical Error}
\label{section:rq3}

RQ3 asks whether models that recover a broad weakness family can also recover the \emph{exact} \gls{CWE} and place it correctly in the taxonomy. We use Top-$k$, MRR, and \gls{HDD} to capture ranking quality and hierarchical error structure.

\begin{figure*}[!t]
    \centering
    \includegraphics[width=\textwidth]{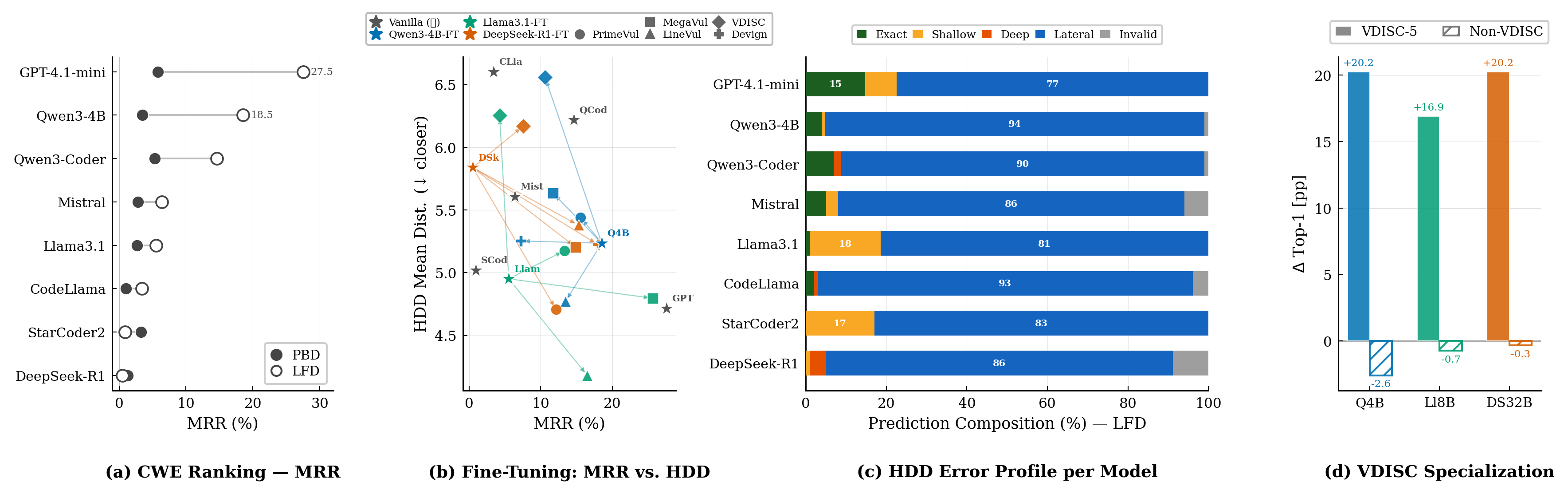}
    \vspace{-12pt}
    \caption{RQ3 semantic-depth analysis. Panel~(a): per-model MRR ranking on PBD and LFD. Panel~(b): MRR \emph{vs.}\ \gls{HDD} mean distance for LFD fine-tuned models; arrows connect each vanilla backbone to its fine-tuned variants. Panel~(c): per-model \gls{HDD} error profile (Exact / Shallow / Deep / Lateral / Invalid). Panel~(d): VDISC-FT specialization on its five labeled \glspl{CWE} \emph{vs.}\ all remaining \gls{CWE} classes.}
    \label{fig:rq3_semantic_depth}
    \vspace{-12pt}
\end{figure*}

\subsection{Finding 1: CWE Exact Identification Is Unsolved}
\label{section:rq3_low_exact}

Figure~\ref{fig:rq3_semantic_depth}(a) shows that exact semantic recovery remains rare. The best vanilla Top-1 on LFD is $14.71\%$ (GPT-4.1-mini), the best MRR is $27.55\%$, and DeepSeek-R1 and StarCoder2 score $0\%$ Top-1. GPT-4.1-mini's MRR rises from $5.79\%$ on PBD to $27.55\%$ on LFD, but much of that increase is explained by LFD's smaller candidate label space (19 CWEs vs.\ 74) rather than a qualitative jump in semantic understanding. Any residual gap may reflect differences in sample difficulty or class composition between the two splits.

\subsection{Finding 2: Lateral Errors Are the Universal Dominant Failure Mode ($86.5\%$ Mean)}
\label{section:rq3_lateral}

Figure~\ref{fig:rq3_semantic_depth}(c) decomposes \gls{HDD} error types on LFD. Across all eight vanilla models, $86.5\%$ of predictions are lateral errors: the model lands at the correct hierarchy depth but in the wrong sibling branch. Only $4.2\%$ are exact matches. Table~\ref{tab:top5-lfd-cwes} shows the same pattern for CWE-416 and CWE-667, confirming that models reach the right specificity level but not the correct sibling category (a structurally different failure mode from shallow or unrelated guessing).

\subsection{Finding 3: MRR and HDD Are Only Weakly Coupled, Validating HDD as a Distinct Diagnostic}
\label{section:rq3_hdd_independent}

Figure~\ref{fig:rq3_semantic_depth}(b) plots LFD MRR against \gls{HDD} mean distance for fine-tuned models. The weak coupling confirms that \gls{HDD} captures information orthogonal to ranking: Qwen3-Coder reaches MRR~$= 14.61\%$ on LFD but has \gls{HDD} mean distance $= 6.22$. At the same time, Llama3.1 has lower MRR ($5.52\%$) but tighter hierarchical proximity (\gls{HDD}~$= 4.95$). A model that ranks the true \gls{CWE} correctly more often is not necessarily making closer hierarchical guesses when it misses. \gls{HDD} therefore provides a distinct diagnostic view of residual error.

\subsection{Finding 4: VDISC Specialization Is Sharp ($+33.3$~pp) but Trades Hierarchical Plausibility for Exact Match}
\label{section:rq3_vdisc}

Figure~\ref{fig:rq3_semantic_depth}(d) provides the clearest specialization result. On the 15 PBD samples within VDISC's five labeled classes, VDISC-FT improves Top-1 by $+33.3$~pp (Qwen3-4B and DeepSeek-R1-32B) and $+26.7$~pp (Llama3.1-8B), but the same models are flat to slightly worse outside those classes; the pattern persists on LFD. The Top-1 gain is paired with a drop in Hierarchical~\%, meaning fine-tuning trades ``plausibly related wrong'' predictions for ``exactly right or unrelated'' ones. RQ3, therefore, shows that models sometimes recover broad families but still fail to identify the exact root cause, and that narrow supervision produces brittle specialization.

The dominant failure mode is lateral error at the correct hierarchy depth rather than shallow wrong-family guesses, confirming that \gls{HDD} captures residual error structure that ranking metrics alone cannot diagnose.

\begin{table}[!b]
    \vspace{-12pt}
    \centering
    \caption{Top 5 LFD \gls{CWE} classes by support. Hier.  combines shallow, deep, and lateral taxonomy-consistent predictions. Lateral is the dominant component in every row.}
    \label{tab:top5-lfd-cwes}
    \vspace{-6pt}
    \scriptsize
    \setlength{\tabcolsep}{4pt}
    \renewcommand{\arraystretch}{1.0}
    \begin{tabular}{@{}lrrrr@{}}
        \toprule
        \textbf{CWE} & \textbf{Exact \%} & \textbf{Hier. \%} & \textbf{Unrel. \%} & \textbf{Support} \\
        \midrule
        CWE-416      & 5.6               & \textbf{88.1}     & 6.3                & 805              \\
        CWE-125      & 23.6              & \textbf{68.6}     & 7.8                & 322              \\
        CWE-476      & 7.6               & \textbf{85.2}     & 7.2                & 290              \\
        CWE-667      & 0.0               & \textbf{91.3}     & 8.7                & 138              \\
        CWE-787      & 2.9               & \textbf{89.9}     & 7.2                & 138              \\
        \bottomrule
    \end{tabular}
\end{table}

\section{Discussion}
\label{section:discussion}

Two results explain why \gls{LLM}-based vulnerability detection remains unreliable. First, each backbone carries a stable directional prior that dominates accuracy. Second, CVE-level contamination does not translate into a detectable advantage in function-level memorization. These findings explain why larger fine-tuning corpora and richer context have not produced reliable detectors: the bottleneck is not more data but the structure and fidelity of the supervision signal. Both the pre-training prior and the quality of the fine-tuning target matter more than dataset size alone, and the same backbone can look acceptable on one aggregate metric while still failing in ways that make it unreliable for real vulnerability triage.

\subsection{The Backbone's Directional Prior Dominates Fine-Tuning}
The near-random accuracy ($\sim50\%$) observed in RQ1 masks extreme, backbone-determined, and stable directional biases. Two failure modes emerge from DFI analysis:
\begin{itemize}
    \item \textbf{Skeptical prior:} General-purpose models such as GPT-4.1-mini strongly prefer ``Safe'' (DFI $-85.5$~pp on PBD, $-89.3$~pp on LFD).
    \item \textbf{Paranoid prior:} Models such as DeepSeek-R1 (DFI $+94.8$~pp on PBD, $+90.8$~pp on LFD) and code-specialized models like CodeLlama (DFI $+67.8$~pp on PBD) flag nearly all complex C code as vulnerable.
\end{itemize}

These priors persist from PBD to LFD, indicating stable backbone behavior rather than contamination effects. Fine-tuning mainly shifts output thresholds without changing the decision policy. This explains the cross-task inversion, in which DeepSeek-R1 is hardest to improve for detection yet gains the most in coarse \gls{CWE} classification: a paranoid model assigns \emph{some} \gls{CWE} to nearly every sample, thereby boosting coarse taxonomy coverage. Qwen3-4B shows the opposite pattern, with large gains driven by overcoming abstention rather than learning transferable semantics. Adding more L2 context does not escape this dynamic, because context can expose the bug only if the decision policy uses it. Raw accuracy alone, therefore, obscures the difference between genuine uncertainty and systematic directional collapse.

One alternative explanation is regression to the mean: DeepSeek-R1 starts from a weak taxonomy baseline, so almost any training signal could produce a large apparent gain. Regression alone, however, does not explain why its improvements exceed those of backbones with stronger baselines. The DFI evidence provides a more direct mechanism: a paranoid backbone labels nearly everything as vulnerable and therefore assigns \emph{some} \gls{CWE} to almost every sample, boosting coarse coverage without solving binary discrimination.

The implication is direct: selecting a backbone is selecting a failure mode. Raw accuracy alone cannot reveal this; DFI is required to distinguish whether near-$50\%$ accuracy reflects genuine uncertainty or two extreme biases that cancel out.

\subsection{Fine-Tuning Decouples Detection from Understanding}
Across all 15 fine-tuned variants, detection and \gls{CWE} understanding move independently. The clearest example is VDISC: its five labeled \gls{CWE} classes produce large exact-Top-1 gains on covered samples ($+26.7$ to $+33.3$~pp), but not on uncovered ones. Fine-tuning, therefore, organizes a narrow supervised label set without building transferable root-cause semantics. The flat covered-versus-uncovered pattern in RQ1 makes the same point from a detection perspective: most gains are policy shifts, not evidence of broader semantic transfer.

This is also consistent with the label-space transfer analysis in RQ1. Fine-tuning does not help most where the labels should matter most; instead, it tends to reorganize the model's response policy globally. Detection and exact semantic diagnosis, therefore, move on partially independent axes.

\subsection{CVE-Level Contamination Does Not Yield a Detectable Advantage}
\label{sec:data_quality}
Prior work flags contamination as a threat to \gls{LLM} evaluation, but here it yields no measurable gain. The key reason is that \emph{CVE-level overlap does not imply function-level memorization}: the vulnerable function evaluated in PBD was absent from training for the verified contaminated CVEs. With label noise, this leaves an estimated 84\% of nominally contaminated training samples without usable memorization signal (Table~\ref{tab:cwe_accuracy}). This does not mean contamination never matters; rather, CVE overlap alone is an inadequate proxy for memorization in context-aware function-level evaluation.

\textbf{Conflicting Ground Truth (Semantic Conflict):}
Automated extraction includes unchanged ``context'' functions from \gls{CVE}-fixing commits. The same \gls{CVE} identifier is therefore paired with a ``Vulnerable'' label for the root-cause function and ``Safe'' labels for other co-modified functions (a 51:1 raw ratio). Even with oversampling, the supervision signal remains contradictory~\cite{ding2024vulnerability}. This weakens both binary and targeted detection because the model encounters mutually inconsistent examples associated with the same CVE.

\textbf{CWE Misclassification:}
As shown in Table~\ref{tab:cwe_accuracy}, about 30\% of contaminated samples carry incorrect \gls{CWE} tags relative to our ground truth. Common errors swap symptoms for root causes or conflate related privilege-management categories. Because PrimeVul, MegaVul, and LineVul are much larger than PBD, this rate is estimated from the overlap subset rather than the entire training corpora. We exclude Devign and VDISC from this analysis because they do not provide comparable CWE labels.

\textbf{Function-Level Distribution Shift:}
Beyond label noise, fine-tuning data and PBD evaluate different units. Vulnerability datasets use commit-level labeling, so every function touched receives a CVE identifier, whereas PBD extracts the specific root-cause function. We verified this mismatch across the contaminated CVEs. In LineVul, all 8 verified contaminated CVEs place the vulnerable function only in the held-out split, so training sees only safe variants. In MegaVul, contaminated CVEs often map to different functions in different subsystems. CVE-level overlap, therefore, does not imply function-level exposure, even when the nominal CVE identifier matches exactly.

\textbf{Summary.}
Of the 281 contaminated samples, three degradation mechanisms apply at different strengths. The weakest---semantic-conflict supervision---is broad: a 51:1 ratio of safe-context to root-cause labels dilutes any individual CVE's signal but does not fully negate it. The two stronger mechanisms eliminate the signal: (i) 88/281 (31.3\%) carry incorrect CWE tags, meaning the targeted-detection objective contradicts ground truth; and (ii) all 8 LineVul-contaminated CVEs place the vulnerable function exclusively in the held-out split, so training never sees a vulnerable example for those CVEs. Taking the union of the two stronger mechanisms (88 wrong-CWE samples and 49 LineVul held-out samples, with partial overlap) leaves at most $281 - (88 + 49 - \text{overlap}) \approx 16\%$ of samples whose CWE label is correct \emph{and} whose vulnerable function appeared in training. The remaining ${\sim}84\%$ are compromised by at least one hard failure. This is sufficient to explain why the empirical seen-CVE advantage remains null across backbone families.

\begin{table}[!t]
    \centering
    \scriptsize
    \setlength{\tabcolsep}{3pt}
    \renewcommand{\arraystretch}{1.05}
    \caption{CWE label accuracy on contaminated samples (vs. PBD).}
    \label{tab:cwe_accuracy}
    \begin{tabular}{lrrrr}
        \toprule
        \textbf{Dataset}  & \textbf{Total} & \textbf{Corr.} & \textbf{Incorr.} & \textbf{Acc.} \\
        \midrule
        PrimeVul          & 88             & 61             & 27               & 69.3          \\
        MegaVul           & 144            & 94             & 46               & 67.1          \\
        LineVul           & 49             & 38             & 11               & 77.6          \\
        \midrule
        \textbf{Combined} & \textbf{281}   & \textbf{193}   & \textbf{84}      & \textbf{68.7} \\
        \bottomrule
    \end{tabular}
    \vspace{-12pt}
\end{table}

The full seen-CVE vs.\ clean-CVE comparison appears in Section \ref{section:rq1_contamination} and Figure~\ref{fig:rq1_mechanisms}(b). No fine-tuned family shows a consistent memorization advantage, confirming that CVE identifier overlap is a weak proxy for memorization exposure: the relevant function, label, and context can all differ from what the model saw during training. This does not imply contamination can never matter; rather, the CVE-overlap proxy is too coarse for function-level, context-aware evaluation.

\subsection{Implications for Reliability}
These findings point to two practical bottlenecks: the absence of function-level ground truth that matches evaluation, and backbone priors that override supervision. A balanced in-domain probe remained near chance on both the training and validation splits, suggesting weak, biased supervision rather than overfitting. A more promising path is higher-fidelity training data that pairs each CVE with its root-cause function, combined with fine-tuning methods that counter directional collapse instead of assuming balanced supervision alone will correct it.

Three concrete directions follow. First, cross-language replication (Java, Python, Rust) would test whether directional priors and threshold-shift findings are specific to low-level C; the framework's language-agnostic pipeline supports this. Second, structured prompting strategies (few-shot, chain-of-thought, retrieval-augmented context) may mitigate the impact of backbone priors without retraining, and CWE-Trace provides the contamination-controlled benchmark needed to measure such interventions. Third, contrastive objectives or DPO-style preference training on paired vulnerable--patched examples are natural candidates for correcting directional collapse, and CWE-Trace's paired structure and DFI diagnostic support this study. We do not claim these alternatives fail; rather, this study characterizes the failure modes that such alternatives should be evaluated against. This shifts the research agenda from scaling noisy labels to building supervision that matches the reasoning unit being evaluated.

\section{Threats to Validity}
\label{section:threatstovalidity}

\textbf{Internal Validity.} To control the non-determinism of \glspl{LLM}, we set the temperature to 0 in all experiments, ensuring reproducibility. We note that temperature~$=0$ may amplify each model's dominant output mode, and non-zero temperatures could in principle reduce DFI magnitude. However, the biases we observe (DFI spanning $-85.5$ to $+94.8$~pp) far exceed plausible stochastic variation and persist across PBD and LFD, indicating intrinsic priors rather than sampling artifacts. Temperature~$=0$ is thus conservative: it makes biases maximally visible rather than masking them.

A primary concern is \emph{data contamination} via post-release updates. We address this with a two-tier provenance verification strategy (Table~\ref{table:llm-specs}): Tier~1 models are pinned to final pre-2025 HuggingFace commit SHAs, and Tier~2 models released in 2025 are pinned to initial public SHAs whose knowledge cutoffs predate LFD compilation. The consistency of DFI patterns across PBD and LFD confirms the absence of performance discontinuities typical of memorization. We also quantify overlap between PBD and common fine-tuning datasets; Section \ref{sec:data_quality} shows this CVE-level overlap yields no detectable advantage because root-cause functions were absent from training. The temporal split is therefore conservative by construction and remains useful despite the null contamination result. We further find that 59.62\% of PBD CVEs appear in at least one fine-tuning dataset (MegaVul 54.72\%, PrimeVul 33.96\%, LineVul 18.11\%), underscoring the need for a leakage-free split.

Unlike prior work that used automated scripts, we apply a two-reviewer manual inspection protocol with consensus to filter out non-security changes and reduce label noise. This is supported by our audit (\S\ref{sec:data_quality}), which found a 75\% label conflict rate in PrimeVul and MegaVul. Manual verification does not remove all subjectivity, but substantially reduces the noise sources identified in \S\ref{sec:data_quality}.
The same protocol addresses a second internal validity threat: incomplete vulnerability context. Because samples come from kernel commits rather than reconstructed environments, manual review ensures retained code still exposes the flaw. Although context-aware blocks include surrounding code, we do not reconstruct full build environments (includes/macros). We mitigate this by verifying that context is sufficient for expert identification, though some vulnerabilities may still depend on environmental details not fully captured. This mainly affects hardware-specific paths and complex conditional compilation regions, which we exclude when context cannot be reliably recovered.

Our supervised transfer study uses three backbones with LoRA adaptation, so findings apply most directly to LoRA-based fine-tuning on these models and datasets; other PEFT methods or full fine-tuning may differ. We treat these as directions for future work rather than contradictions. Similarly, we study five widely used vulnerability datasets rather than exhaustively covering all academic corpora, so results primarily reflect current mainstream fine-tuning practice.

The evaluation also uses zero-shot prompts. This choice isolates pre-training and fine-tuning effects from prompt engineering and matches a triage setting where no per-vulnerability demonstrations are available, but it does not rule out gains from few-shot prompts, chain-of-thought prompting, retrieval-augmented context, tool use, or agentic workflows. Our conclusions therefore concern the zero-shot and supervised-LoRA regimes evaluated here.

\textbf{External Validity.} Our study focuses on the Linux kernel (C language), a deliberate design choice: the kernel is among the most complex and security-critical real-world codebases, with concurrency, resource-lifetime, and privilege-management vulnerabilities requiring multi-file reasoning. If LLMs fail here, that failure is informative precisely because the kernel represents a demanding upper bound for systems software. While this limits direct claims about other languages, the failure modes we identify (stable directional priors and threshold-shifting fine-tuning) are architectural in nature and not inherently tied to C or the kernel's coding idioms. We therefore expect them to manifest in any domain where fine-tuning supervision is noisy and context-aware evaluation is applied. The magnitude of the observed effects may differ in domains with shorter functions, simpler control flow, or weaker dependence on cross-file state.

\textbf{Construct Validity.} Binary detection and CWE ranking do not fully capture internal reasoning. We mitigate this by adding HDD for semantic proximity and error direction, DFI for directional bias, and abstention as a valid triage outcome. These measures do not reveal internal reasoning directly, but they provide a more discriminating behavioral picture than exact-match accuracy alone. HDD is fully determined by the CWE-1000 taxonomy graph and contains no subjective weight parameters, addressing subjectivity concerns raised for earlier hierarchical metrics such as HPS. Results are tied to the CWE-1000 version used in this study, but are stable within any consistent version of that taxonomy.

\section{Conclusion}
\label{section:conclusion}

CWE-Trace stress-tests the zero-shot security behavior of \glspl{LLM} on 834 real-world Linux kernel samples using temporal splits, context-aware extraction, and hierarchy-aware evaluation. The framework's three core components are reusable: the extraction pipeline is language-agnostic, HDD requires only the CWE-1000 taxonomy graph and no subjective weights, and DFI can be computed from any four-task detection experiment.

Two results stand out. First, CVE-level contamination does not yield a detectable advantage: verified overlaps either exclude the vulnerable function from training or map the CVE to unrelated functions, and combined with semantic-conflict labels and 31\% CWE misclassification, an estimated 84\% of nominally contaminated samples carry no usable memorization signal.
Second, backbone directional priors dominate fine-tuning. Models retain stable paranoid or skeptical failure modes (DFI spans $-85.5$ to $+94.8$~pp) from historical to post-cutoff data, and fine-tuning mainly shifts the output threshold rather than the underlying decision policy. The path to reliable AI-assisted security lies not in larger context windows but in better-structured supervision, prior-correcting training objectives, and evaluation frameworks that distinguish genuine reasoning from threshold-shifting.

Future work should prioritize corpora that pair each CVE with its root-cause function, extend evaluation to other languages and ecosystems, and explore contrastive training objectives that force models to learn vulnerability semantics rather than superficial correlations.

\bibliographystyle{IEEEtran}
\bibliography{ref}

\end{document}